\documentclass[useAMS,usenatbib]{mn2e}
\usepackage{times,epsfig,natbib,amssymb,amsmath,graphics,longtable,threeparttable}

\def \lsun{\ifmmode{{\rm\ L}_\odot}\else{${\rm\ L}_\odot $}\fi}
\def \msun{\ifmmode{{\rm\ M}_\odot}\else{${\rm\ M}_\odot$}\fi}
\def \rsun{\ifmmode{{\rm\ R}_\odot}\else{${\rm\ R}_\odot$}\fi}
\newcommand{\kms}{kms$^{-1}$}                         
\def \mdot{\ifmmode{{\rm\dot{M}}}\else{${\rm\dot{M}}$}\fi}

\newcommand\as{${''}$}

\newcommand{\apj}{ApJ}
\newcommand{\apjl}{ApJ Let.}
\newcommand{\aj}{AJ}

\newcommand{\aap}{A\&A}

\newcommand{\mnras}{MNRAS}

\newcommand{\araa}{ARA\&A}

\newcommand{\nat}{Nature}

\newcommand{\pasp}{PASP}

\newcommand{\iaucirc}{IAUC CBET}

\newcommand{\R}{\textit{R}}

\newcommand{\ha}{H$\alpha${}}

\newcommand{\NII}{[N{\sc ii}]}                  

\title[Constraints on CC SN progenitors]{Constraints on core-collapse supernova progenitors from correlations 
with \ha\ emission\thanks{Based 
on observations made with the Isaac Newton Telescope operated on the island
of La Palma by the Isaac Newton Group in the Spanish Observatorio del Roque de los
Muchachos of the Institute de Astrofisica de Canarias, and 
on observations made with the Liverpool Telescope operated on the island of La Palma by Liverpool John Moores 
University in the Spanish Observatorio del Roque de los Muchachos of the Instituto de Astrofisica 
de Canarias with financial support from the UK Science and Technology Facilities Council.}}
\author[J.P. Anderson and P.A. James]{J.P. Anderson\thanks{E-mail:jxa@astro.livjm.ac.uk} and P.A. James\\
Astrophysics Research Institute, Liverpool John Moores University,
Twelve Quays House, Egerton Wharf, Birkenhead CH41 1LD, UK}
\begin{document}

\date{accepted 2008 August 14}

\pagerange{\pageref{firstpage}--\pageref{lastpage}} \pubyear{2008}

\maketitle

\label{firstpage}

\begin{abstract}
We present observational constraints on the nature of the different core-collapse supernova types
through an
investigation of the association of their explosion sites
with recent star formation, as
traced by \ha +\NII\ line emission.
We discuss results on the analysed data of the positions of 168 core-collapse
supernovae with respect to the \ha\ emission within their host galaxies.\\
From our analysis we find that overall the type II progenitor population 
does not trace the underlying star formation. Our results are consistent
with a significant fraction of SNII arising from progenitor stars of
less than 10\msun . 
We find that the supernovae of type Ib 
show a higher degree of association with HII regions 
than those of type II (without accurately tracing the emission), 
while the type Ic population accurately traces the \ha\ emission.
This implies that the main core-collapse supernova types form a sequence of 
increasing progenitor mass, from the type II, to Ib and finally Ic. 
We find that the type IIn sub-class display a similar 
degree of association with the line emission to the overall SNII population, implying
that at least the majority of these SNe do not arise from the most massive stars.
We also find that the small number of SN `impostors' within our sample do
not trace the star formation of their host galaxies, a result that would not be expected if 
these events arise from massive Luminous Blue Variable star progenitors.
\end{abstract}


\section{Introduction}
\label{intro}
Despite years of observational and theoretical research on the nature of supernova (SN) explosions and the properties
of their progenitors there remain substantial gaps in our knowledge of all SN types. Although
there are many different theoretical predictions as to the nature of SN progenitors,
the observational evidence to discriminate between
various progenitor scenarios remains sparse. 
SNe can be split into two theoretical classes; SNIa which are thought
to arise from the thermonuclear explosion of an accreting white dwarf, and 
core-collapse (CC) SNe which are believed to signal the collapse of the cores of massive 
stars at the end points in their stellar evolution. \\
Results from the first paper in this series (\citealt{jam06}, JA06 henceforth) suggested that the
SNIb/c arise from higher mass progenitors than SNII (albeit with small statistics; only 8 SNIb/c).
We test these initial results with an increased sample size enabling us to distinguish
between the various CC sub-types, and present results from a combined sample of 100 SNII
(that can be further separated into 37 IIP, 8 IIL, 4 IIb, 12 IIn), 62 SNIb/c
(22 Ib, 34 Ic and 6 that only have Ib/c classification), and 6 SN `impostors'.
We will present
results and discussion of SNIa within the context of the methods used in this
paper elsewhere. We will also present further research on the radial 
positions of SNe within galaxies, and on correlations between CC SN type and local metallicity
in future publications. Here we concentrate on results on the progenitor masses
of the different CC SNe.

\subsection{Core-collapse supernovae}
\label{cc}
CC SNe are thought to be the final stage in the stellar evolution
of stars with initial masses $>$8-10\msun , when fusion ceases in the cores
of their progenitors and they can no longer support themselves against 
gravitational collapse. The different types of CC SNe are classified according to 
the presence/absence of spectral lines in their early time spectra, plus
the shape of their light curves. The first major classification comes from the presence of
strong hydrogen (H) emission in the SNII. SNIb and Ic lack any detectable
H emission, while the SNIc also lack the helium lines seen in SNIb. SNII
can also be separated in various sub-types. SNIIP and IIL are classified in terms of
the decline shape of their light curves (\citealt{bar79}; plateau in the former and linear in the latter),
thought to indicate different masses of their H envelopes prior to SN,
while SNIIn show narrow emission lines within their spectra \citep{sch90}, thought to arise
from interaction of the SN ejecta with a slow-moving circumstellar medium (e.g. \citealt{chug94}).
SNIIb are thought
to be intermediate objects between the SNII and Ib as at early times their
spectra are similar to SNII (prominent H lines), while at later times they
appear similar to SNIb (\citealt{fil93}).\\
Strong evidence has been presented to support the belief that SNII and SNIb/c arise from massive progenitors, 
through their
absence in early type galaxies \citep{bergh02}, and the direct detection 
of a small sample of progenitors on pre-explosion images (\citealt{sma04,maun05,hen06,li06,gal07,li07,crock08}). However, it is unclear
how differences in the nature of their progenitors produce the
different SNe we see. It is clear that there must be some process by which the progenitors
of the different SNe lose part (or almost all in the case of SNIb and Ic) of their envelopes
prior to explosion. The differences in efficiency of this mass loss process could be 
dependent primarily on progenitor mass, with higher mass progenitors having
higher mass loss rates due to stronger stellar winds, and losing more of their envelopes. In this picture a sequence
of SNe types emerges from SNIIP and IIL to SNIIb, SNIb and finally Ic having successively higher 
initial masses. There are also other factors that probably play an important role. Initial chemical
abundance will also affect the progenitor mass loss, with higher metallicity producing
stronger radiatively driven winds ({e.g. \citealt{pul96,kud00,mok07}).
It has also been proposed \citep{pod92} that massive binaries could
produce a significant fraction of CC SNe, with mass transfer ejecting matter
and leading to some of the various CC sub-types.\\
Since the theoretical separation of SNe into two distinct explosion classes by \cite{hoy60},
there have been many predictions as to how the different CC SN types emerge from
different progenitors. There are two main theoretical routes to achieving the observed different
SN types. The first attempts to describe the full range of SNe from a single star
progenitor scenario. \cite{heg03} and \cite{eld04} produced SN progenitor maps showing how
variations in initial mass and metallicity produced the different CC SN types. These models both
predict that single stars of up to $\sim$25-30\msun\ will produce SNIIP, with stars of slightly 
higher mass producing SNIIL and IIb, and those of $>$30\msun\ ending their lives as SNIb/c (both
authors also predict that these initial mass ranges will shift to higher values with 
decreasing metallicity). In both of these models no attempt was made to differentiate between
the SNIb and the SNIc, but one would presume that within this single star scenario SNIc would arise 
from higher mass progenitors than the SNIb as they have lost even more of their stellar envelopes.    
Alternatively it could be that massive binaries produce the majority of CC SNe other
than SNIIP (with these SNe still arising from single star progenitors). The initial mass of
the stars producing SNIb/c, SNIIL and SNIIb would then be similar to those of SNIIP (12-20\msun , e.g. \citealt{shi90}) but would arise from
binary evolutionary processes. There is also
a growing number of SNe that show evidence of binarity (e.g. SN 1987A; \citealt{pod90} and SN 1993J; \citealt{nom93,pod93,maun04}).
Recent comparisons of the observed ratio of SNIb/c rates to those of SNII also argue that
binaries are playing the dominant role in producing SNIb/c \citep{kob07}, while 
\cite{eld08} predict a SNIb/c rate produced by a combination of single and binary progenitors that
best produces the observed SN rate. Again one should note that these binary models group
SNIb and Ic together and do not attempt to predict what differences in progenitor produce these two types.\\
Given the different predictions for the origin of the CC SN types described above, 
observations are needed to discriminate between these models and thus firmly tie
down the progenitors of the different SN types. However, apart from a small number of
direct detections of progenitors (\citealt{sma04,maun05,hen06,li06,gal07,li07,crock08}) this observational evidence remains sparse. 
Therefore here we present results to test the above predictions and constrain
differences in progenitor mass of the different CC SN sub-types by investigating the nature of their parent 
stellar populations within host galaxies.
 
\subsection{Progenitor constraints from parent stellar populations}
\label{pops}
The most obvious way to determine the nature of SN progenitors is to 
investigate the properties of their stars on pre-explosion images. This has had some 
success although
it is only possible for events in very nearby galaxies and therefore the
statistics remain low. Another way is to investigate how the rates of the various
SN types vary with different parameters, such as redshift or host galaxy properties.
Our approach is intermediate to these methods as we attempt to constrain the nature 
of SN progenitors through investigating the environments and stellar 
populations at the positions of historical SNe. Here we concentrate on the association
of the different CC SNe types with recent star formation (SF) as traced by
\ha\ emission.\\
\cite{ken98} states in a review paper on \ha\ imaging techniques that:
``only stars with masses $>$10\msun\ and lifetimes of $<$20 Myr contribute significantly 
to the ionising flux''. Thus, if our understanding of this line emission is correct,
we can use this assumption as a starting point to constrain the relative stellar
lifetimes and therefore the relative masses of the various SN progenitors, through
investigating how accurately the different SN types trace the emission.
In JA06 we 
presented a statistic to quantitatively measure the association of individual
SNe with the \ha\ emission of their host galaxies, and presented
results from an initial galaxy sample (\ha GS, discussed in \S \ref{data}).
It was found that overall the SNII progenitor population did not trace the 
underlying SF of their host galaxies, with a significant fraction lying 
on regions of low or zero emission line flux which were ascribed to
a `Runaway' fraction of progenitor stars
(however, this assumed that SNII arise from progenitors of $>$10\msun ).
The SNIb/c did appear to 
follow the emission implying that these progenitors come from higher mass
stars than the SNII, although the statistics on this class were small (only 8 SNe 
for SNIb and Ic combined). This SN/galaxy sample has now been significantly increased,
enabling the full parameter space of CC SN progenitors to be investigated and results 
from this increased sample are presented here.\\
The paper is arranged as follows: in the next section we present the data and 
discuss the reduction techniques employed, in \S \ref{ncr} we summarise the 
statistic introduced in JA06 and used throughout this paper, in \S \ref{results}
we present the results for the different CC SN types, in \S \ref{diss} we discuss
possible explanations for these results and their implications for the relative masses 
of the SN progenitors, and finally in \S \ref{con} we draw our conclusions. 

\section{Data}
\label{data}
The initial galaxy sample that formed the data set for JA06 was the \ha\ Galaxy Survey (\ha GS).
This survey was a study of the SF properties of the local Universe using \ha\ imaging of
a representative sample of nearby galaxies, details of which can be found in \cite{jam04}.
63 SNe (of all types, including SNIa) were found to have occurred in the 327 \ha GS galaxies
through searching the International Astronomical Union (IAU) database
\footnote{http://cfa-www.harvard.edu/iau/lists/Supernovae.html}.\\
Through three observing 
runs on the Isaac Newton Telescope (INT) and an on-going time allocation with
the Liverpool Telescope (LT) we have now obtained \ha\ imaging for the host galaxies of
133 additional CC SNe, the analysis of which is presented here. The LT is a fully robotic 2m telescope
operated remotely by Liverpool John Moores University. To obtain our imaging we used 
\textit{RATcam} together with the narrow \ha\ and the broad-band Sloan \textit{r'} filters. Images were binned 
2$\times$2
to give us 0.278\as\ size pixels, and the width of the \ha\ filter enabled us to image
target galaxies out to $\sim$2400 \kms. The INT observations used the Wide Field Camera (WFC) together
with the Harris \textit{R}-band filter, plus the rest frame narrow \ha\ (filter 197)
and the redshifted \ha\ (227) filters enabling us to image host galaxies out to $\sim$6000 \kms. During our
2005 INT observing run we also used the SII filter (212) as a redshifted \ha\ filter and imaged 12 
SN hosting galaxies at distances of $\sim$7500 \kms. The pixel scale on all INT images is 0.333\as\ per pixel
and with both the LT and INT our exposure times were $\sim$800 sec in \ha\ and $\sim$300 sec in \textit{R}.\\ 
These additional SNe/galaxies 
were chosen from the Padova-Asiago SN catalogue\footnote{http://web.pd.astro.it/supern/},
as specific CC SN types were more complete for the listed SNe. 
At a later date all SN type classifications taken from the Padova-Asiago catalogue
were checked through a thorough search of the literature and IAU circulars, as classifications
can often change after the initial discovery and therefore those in the catalogue may not be completely accurate.
The full list of SN types is given in Appendix B, where references are given if classifications were changed from
those in the above catalogue. The main discrepancies were the classification of the so called SN `impostors'
as SNIIn in the Padova-Asiago catalogue. These are transient objects that are believed to be the outbursts from very massive Luminous Blue Variable
stars (LBVs), which do not fully destroy the progenitor star and are therefore not classed as true SNe (e.g. \citealt{van00,maun06}).
Six such objects were found in our sample, and the results on these `impostors' are presented and discussed separately
in the following sections.\\
The distance limit for our sample (mainly set
from the available \ha\ filters during observing runs) enables us
to resolve the stellar population close to the SN position, and we also exclude edge on galaxies
because of extinction effects and increased projection uncertainties. 
We do not include results on SNe where images were obtained within 18 months for SNII and a year for SNIb/c
after the catalogued explosion epoch. This is to ensure that our images are not contaminated with residual SN light and that the \ha\ emission that we detect
is due to the underlying HII regions and not associated with the SNe themselves.
Through the above telescope
time allocations we have therefore obtained data on host galaxies of almost all 
discovered CC SNe (that have been classified IIP, IIL, IIb, IIn, Ib, and Ic) that meet our
selection criteria and were observable within the \ha\ filters of the two telescopes.\\
There are obvious biases within a set of data chosen in the above way. As we use 
any discovered SNe for our sample, the various different biases in the different SN surveys
that discovered them mean that the galaxy/SN sample is by no means representative of the 
overall SN populations. Bright, well studied galaxies will be over represented, as will brighter
SNe events that are more easily detectable. However, firstly we are not analysing the overall host 
galaxy properties (as we will show when discussing the statistics we use in \S \ref{ncr}), but
are analysing where within the distribution of stellar populations of the host galaxy the SNe are occurring. 
Secondly, the small number of CC sub-types that are discovered means that 
no individual survey can currently manage to analyse the properties of their host galaxies
or parent stellar populations in any statistically significant way 
(most statistical observational studies do not even 
attempt to separate the Ib and Ic SN types). Taking our approach enables us to make statistical
constraints on all the major CC SN sub-types. The results that are presented 
in this paper are on the analysis of the parent stellar populations of 100 SNII, of which 37 are IIP,
8 IIL, 4 IIb and 12 IIn, 6 SN `impostors', plus 22 Ib, 34 Ic and 6 that only have Ib/c as their classification, from both
the initial \ha GS sample and our additional data described above.

\subsection{Data reduction and astrometric methods}
\label{reduce}
For each SN host galaxy we obtained \ha +\NII\ narrow band imaging, plus \textit{R}- or \textit{r'}-band imaging
used for continuum subtraction. Standard data reduction (flat-fields, bias subtraction etc) were 
achieved through the automated pipeline of the LT \citep{ste04}, and the 
INT data were processed through the INT Wide Field Camera (WFC), Cambridge Astronomical Survey Unit (CASU)
reduction pipeline. Continuum subtraction
was then achieved by scaling the broad-band image fluxes to those of the \ha\ images using 
stars matched on each image, and subtracting the broad-band image from the \ha\ images. Our reduction
made use of various \textit{Starlink} packages.\\
The next process was to obtain accurate positions for the sites of our SNe on their host
galaxy images. This astrometric calibration was achieved by transferring the accurate
astrometry of XDSS second generation Palomar Sky Survey images\footnote{downloaded from http://cadcwww.dao.nrc.ca/cadcbin/getdss},
onto matching galaxy images in our sample (the full process is described in JA06).
In nearly all cases astrometric calibration was achieved with fit residuals of $<$0.2\as .
With accurate positions obtained for the SNe sites we 
could now analyse to what degree the different SNe were associated with the distribution of \ha\ emission within their 
host galaxies.\\
In figures~\ref{2001ac} and~\ref{2004bm} we show two examples of \ha\ images of the host galaxies of SNe from our sample,
with SN positions derived from the above astrometric calibration.
We intend to present all our \ha\ and \R -band imaging of SN host galaxies in a future publication (Ivory et al. 2008, 
in prep), where we will release all of our data for public use along with host galaxy derived characteristics such
as SF rates and \ha\ equivalent widths.

\begin{figure}
\includegraphics[width=8.5cm]{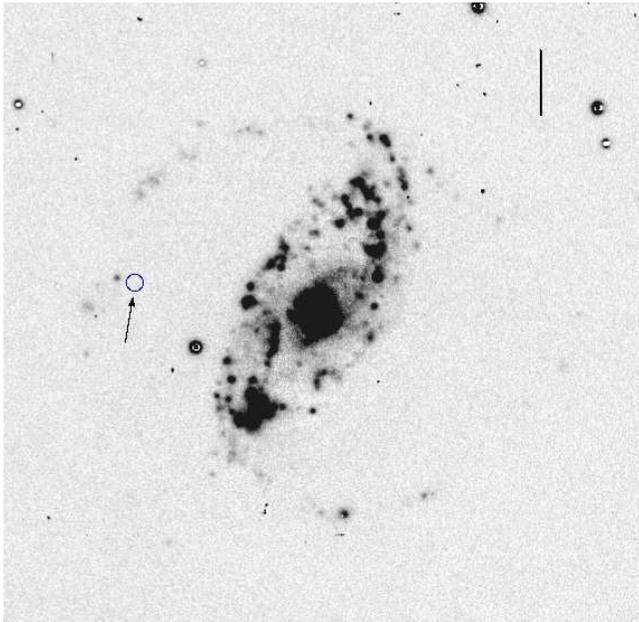}
\caption{An example negative continuum subtracted \ha\ image from our data (image taken with the WFC on the INT); SN 2001ac (SN `impostor') 
(position indicated by circle/arrow), within
the host galaxy NGC 3504. The scale bar is 20\as . The NCR value for this SN is 0.000.}
\label{2001ac}
\end{figure}

\begin{figure}
\includegraphics[width=8.5cm]{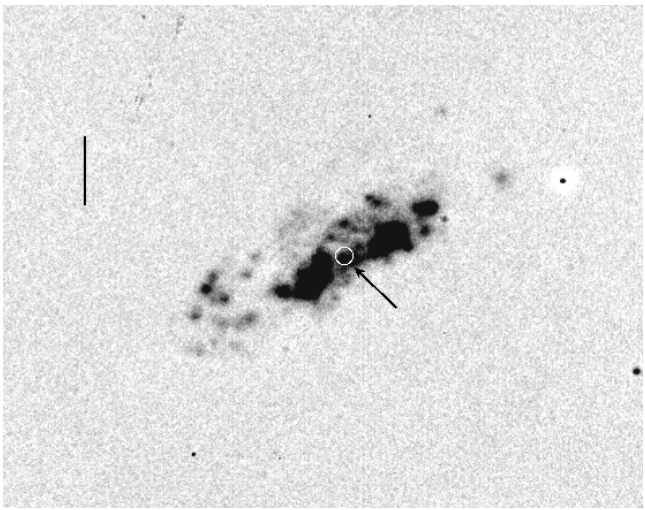}
\caption{Another example negative continuum subtracted \ha\ image (WFC, INT); SN 2004bm (Ic) (position indicated by circle/arrow), within
the host galaxy NGC 3437. The scale bar is 20\as . The NCR value for this SN is 0.704.}
\label{2004bm}
\end{figure}

\section{Pixel statistics}
\label{ncr}
Previous works investigating the association of SNe with HII regions within host galaxies (e.g. \citealt{bart94,vandyk96})
have generally used some measure of the distance to the nearest bright HII region to
quantify the association of each individual SN to the high mass SF within their host galaxies.
However, this brings various problems when defining the nearest bright HII region 
and therefore the distance to measure. In JA06 we presented a quantitative statistic that reduced any
ambiguity in the measurements of each SN, by analysing where the count of the SN hosting 
pixel falls within the overall distribution of \ha\ pixel values of the galaxy. The exact details
of how this statistic is formed can be found in JA06, and here we summarise this process
and the main points on how this can be used in analysing the associations of the 
different SN types to the emission.\\
The pixels in the continuum subtracted \ha\ images were first binned 3$\times$3 to reduce the
pixel-to-pixel noise level and enable us to determine the SN-containing pixel with a degree 
of certainty. The pixels were then sorted into increasing pixel count. The cumulative 
distribution of these values was then formed and normalised, with negative values set
to zero, giving a normalised cumulative rank pixel value function (NCR henceforth) running
from 0 to 1, with one entry for each pixel on the host galaxy image. Within
this distribution therefore, values of 0 correspond to zero emission line flux or sky values,
whereas a value of 1 corresponds to the centre of the brightest HII region on the image.
Figures~\ref{2001ac} and~\ref{2004bm} illustrate the use of this statistic, with the SN `impostor' 2001ac falling away
from any detected \ha\ emission in Fig.~\ref{2001ac} and therefore having an NCR value of 0.000, whereas
in Fig.~\ref{2004bm} the SNIc 2004bm falls on a bright HII region and therefore has a high 
NCR value of 0.704.\\
When we form the NCR it is found that the majority of values lying above the sky level within this distribution
are small and individually contribute little to the overall flux, but by force of numbers they
do contribute a significant amount to the underlying SF. Alongside this, there will
be relatively few NCR values close to 1, but those that are will individually
make a significant contribution to the overall flux. Thus the distribution
is formed so that if a SN progenitor population is drawn from the same stellar
population that produces the \ha\ flux, one would expect a mean NCR
value for that SN type of 0.5 and a flat distribution. This is therefore the initial
hypothesis that we work from, that if the progenitors of CC SNe trace the same high mass
SF as does \ha\ emission, we expect their NCR values to form a flat distribution. 
We can then investigate whether there are any differences in the mean NCR values and distributions
of the different CC SN sub-types and what this may imply for differences in the relative lifetimes and masses of their progenitors.\\
A full discussion of the errors associated with this statistic was presented in JA06, therefore here 
we will summarise the main errors; those presented with the results are the
statistical errors found on the various distributions. The most obvious error is that
associated with the determination of the SN containing pixel. This was investigated 
by determining the NCR value of each SN for a 3$\times$3 pixel box  centred on the SN pixel (meaning that after already binning 3$\times$3 
we sample regions $\sim$2.5\as\ and 3\as\ on the LT and INT images respectively).
A comparison was then made of the median NCR value of the box with the SN pixel. This was repeated for the new 
sample where we find the size of the errors to be consistent with those from JA06,
and there are
in general no significant differences between the SN pixel NCR values and those 
of the median value of the surrounding pixels. For the overall SNII NCR distribution
we find a mean difference of 0.027 between the NCR value of the SN pixel and the median pixel. 
The rms difference in NCR value is 0.163 where, as in JA06 this is dominated by around five cases where
there is a significant difference between the values. However, overall the NCR analysis seems
to give results which are robust to positional errors of 1-2\as . In JA06 possible errors due to
the adopted sky level were investigated but these were found to be insignificant. Finally
a Monte Carlo analysis was performed on the effects of pixel-to-pixel noise on the NCR value.
Again this effect was found to be small, with errors appearing to be random and producing no tendency
to bias the results in any particular direction. 
We will now present the results formed from using the above described statistic on the 
various CC SN types.

\section{Results}
\label{results}

\subsection{SNII}
\label{II}
Figure~\ref{figII} shows the overall distribution of NCR values for the 100 SNII our sample. It 
is immediately clear that the positions of SNII do not follow the overall distribution of SF as traced by the \ha\ line
emission, confirming the result of JA06. In fact there is an excess of $\sim$35\%\ of SNII that
fall on sites of little or zero \ha\ flux compared to what would be expected if these SNe followed the distribution of \ha\ emission. 
The probability of the SNII progenitor population being drawn from a flat 
distribution (i.e. following the line emission), calculated using a Kolmogorov-Smirnov (KS) test
is $<$1\%. Overall the mean NCR value for SNII is 0.252 with a standard 
error on the mean of 0.027. We will now present the results obtained when separating the SNII
into their various sub-types. It should be noted here that $\sim$40\%\ of our type II SNe do not have
designated sub-types and are only classified as SNII.

\begin{figure}
\includegraphics[width=9cm]{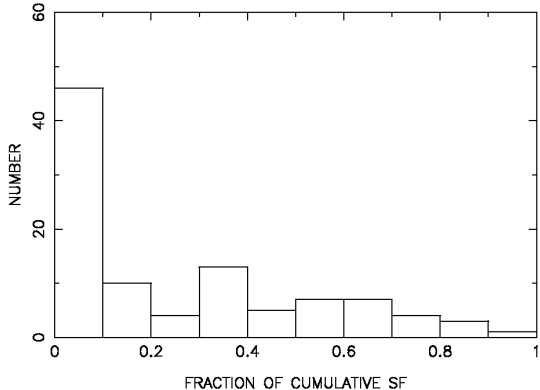}
\caption{Histogram of the NCR values of the SN-containing pixels within the cumulative \ha\ flux
distribution for the 100 SNII}
\label{figII}
\end{figure}

\subsubsection{SNIIP}
\label{IIP}
SNIIP are the most abundant SNII sub-type observed and therefore it is not surprising that these are the most
abundant of those with sub-type classification in our sample. It is also to be expected that their distribution of NCR values
follows that of the overall SNII population as can be seen when comparing Figs.~\ref{figII} and~\ref{figIIP}, 
with a KS test showing that
the two distributions (SNe classified as IIP removed from the overall II distribution) are formally 
consistent with each other. Again, if one assumes that the majority of those unclassified SNII will 
be of type IIP (i.e. if sufficient data were available on their light curves etc), this is to be expected.
The mean NCR value for the SNIIP population is 0.263 (0.048).

\begin{figure}
\includegraphics[width=9cm]{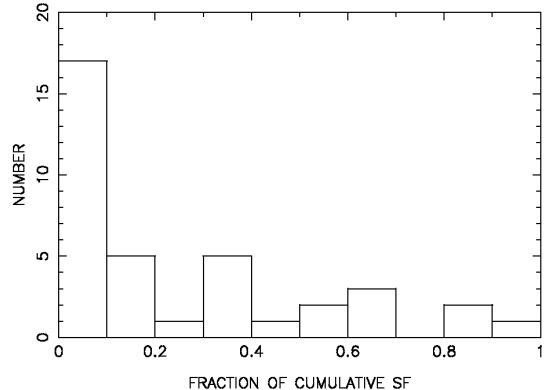}
\caption{Histogram of the NCR values of the SN-containing pixels within the cumulative \ha\ flux
distribution for the 37 SNIIP}
\label{figIIP}
\end{figure}

\subsubsection{SNIIL}
\label{IIL}
The 8 SNIIL population have a mean NCR value of 0.255 (0.112) and seem to follow the same distribution
as the overall SNII population.

\subsubsection{SNIIb}
\label{IIb}
The 4 SNIIb have a mean NCR value of 0.460 (0.162), higher than that of the overall SNII population.
To measure the significance of this difference we used a Monte Carlo analysis. Removing the SNIIb from 
the distribution of SNII NCR values we calculated the fraction of times that a mean NCR value of $\geq$0.460 (SNIIb 
mean value) was produced when four values were drawn at random from the overall SNII distribution. We
found that there is only a $\sim$6\% 
chance that the SNIIb parent population is drawn from the same distribution as that of the
rest of the SNII.

\subsubsection{SNIIn}
\label{IIn}
Figure~\ref{figIIn} shows the distribution of the NCR values for the 12 SNIIn. The mean NCR value
for this SN type is 0.256 (0.088), and these SNe seem to follow the same stellar population
as that of the overall SNII population. Using a KS test we find that there is only $\sim$1\%\ chance that these SNe
are drawn from a flat distribution (i.e. following the distribution of \ha\ emission). 

\begin{figure}
\includegraphics[width=9cm]{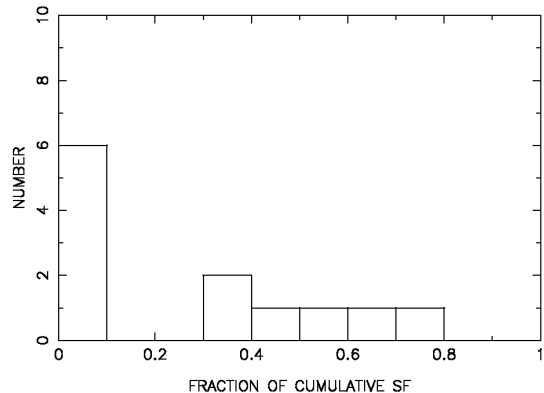}
\caption{Histogram of the NCR values of the SN-containing pixels within the cumulative \ha\ flux
distribution of the 12 SNIIn}
\label{figIIn}
\end{figure}

\subsection{SNIb/c}
\label{Ib/c}
The distribution of NCR values for the 62 SNIb/c is plotted in Fig.~\ref{figbc}. Overall the mean 
NCR value of these SNe is 0.421 (0.040) and these SNe are formally consistent with being drawn from the 
same distribution as that traced by the \ha\ emission, while there is $<$1\% chance that they arise from the
same parent distribution as the SNII. We have presented the results for this overall SNIb/c group to 
make comparisons to the overall SNII progenitor population (as is often quoted elsewhere), however it is clear that in fact the results
for each separate group (Ib, Ic) differ as we will now discuss.

\begin{figure}
\includegraphics[width=9cm]{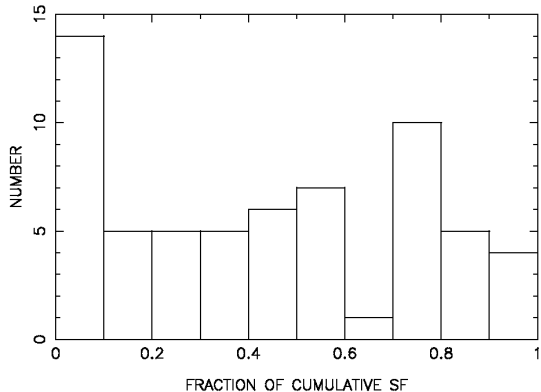}
\caption{Histogram of the NCR values of the SN-containing pixels within the cumulative \ha\ flux
distribution for the 62 SNIb/c}
\label{figbc}
\end{figure}

\subsubsection{SNIb}
\label{Ib}
Figure~\ref{figb} shows the distribution of NCR values for the SNIb population; this SNe type has a mean 
NCR value of 0.367 (0.063). The probability of this SN class being drawn from a flat distribution is $>$10\%. We compare 
this population with that of the SNII and find that although the mean NCR value for the SNIb is higher than that of the SNII,
using a KS test they are formally consistent with being drawn from the same progenitor population ($>$10\%\ chance that they 
arise from the same distribution).

\begin{figure}
\includegraphics[width=9cm]{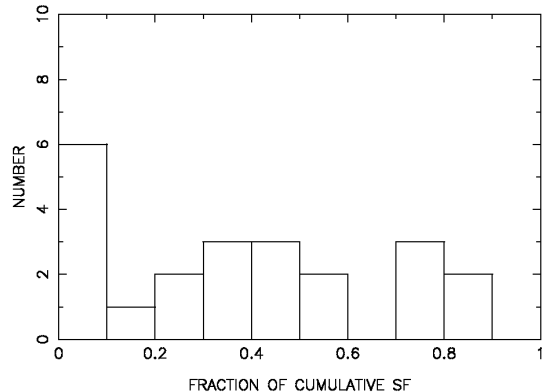}
\caption{Histogram of the NCR values of the SN-containing pixels within the cumulative \ha\ flux
distribution of the 21 SNIb}
\label{figb}
\end{figure}

\subsubsection{SNIc}
\label{Ic}
The distribution of NCR values for the SNIc is shown in Fig.~\ref{figc}. This is the SN type that shows the highest 
degree of association to the recent SF in host galaxies, as traced by \ha\ emission and the 
population has a mean NCR of 0.447 (0.057). A KS test shows that these SNe are formally consistent with being drawn from 
a flat distribution, but there still seems to be a slight excess at zero NCR values. When compared
to the overall SNII distribution we find a $<$1\%\ chance that they are drawn from the same distribution. When
we compare these SNe to the SNIb we find that they have a significantly higher mean value, however there
is still a $>$10\%\ chance that they are drawn from the same parent distribution.\\

\begin{figure}
\includegraphics[width=9cm]{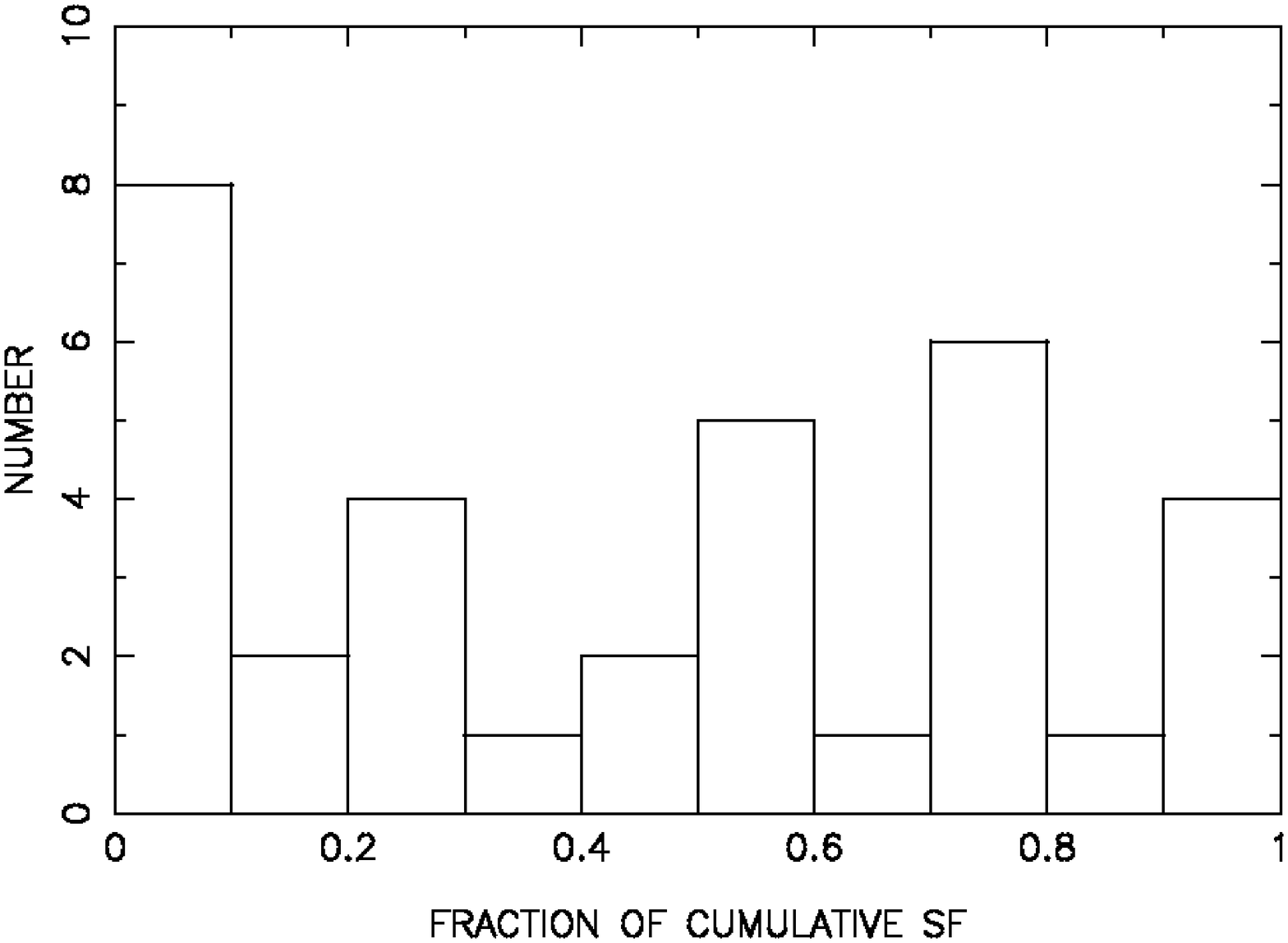}
\caption{Histogram of the NCR values of the SN-containing pixels within the cumulative \ha\ flux
distribution of the 31 SNIc}
\label{figc}
\end{figure}

\subsection{SN `impostors'}
\label{imposres}
The mean NCR value for the 6 SN `impostors' is 0.105 (0.065), considerably lower than that of the SNII
population. To measure the significance of this difference we used a Monte-Carlo analysis as for the SNIIb. We calculated the fraction of 
times that a mean NCR value of $\leq$0.105 (SN `impostors' 
mean value) was produced when six values were drawn at random from the overall SNII NCR distribution. We
found that there is a $\sim$10\%\ chance that the SN `impostors' are drawn from the same distribution as that of
the SNII.

\section{Discussion}
\label{diss}
There are two main discussion points that arise from the above results.
The first is that there is a real excess seen in the number of SNII that
do not appear to show any association to the \ha\ emission, a result
that was seen in JA06 and is backed up with the improved statistics presented within this paper. 
The second is the implications that
differences in NCR values and distributions of the various CC SN types have on 
differences between their progenitor masses. If we assume that all stars originate within 
HII regions (the highest mass stars formed from an episode of SF will start to ionise the local 
hydrogen straight away), 
then the degree of association of each SN type with the overall emission can be
used to constrain their relative stellar lifetimes (and therefore their relative initial masses), as with time, the
stars will either move away from the host HII region due to their peculiar velocities, or
the host HII region will cease to exist as the massive ionising stars will explode as
the first set of SNe. Therefore we discuss the implications of our results for the 
different masses of the different CC types and how these implications fit with other results on the nature of the different SN progenitors.

\subsection{An excess of SNII from regions of zero \ha\ emission}
\label{zeroha}
The results presented in \S ~\ref{II} indicate that around $\sim$35\%\ of SNII fall on sites of
little or zero \ha\ flux, compared to what would be expected if
these SNe followed the underlying SF. For the SNIIP where we have 37 events in our sample this fractional excess remains the same.
Recent research combining the results of a ten year survey for direct detections of SN progenitors (Smartt et al. 2008, in preparation; private communication)
gives additional support to the growing evidence that CC SNe (SNIIP in particular) can arise from stars with initial
masses of less than 10\msun . One of the main results from this survey is a lower mass value for producing SNIIP of 8.5\msun .
Using the initial assumption for the current research that only stars with masses $>$10\msun\ contribute significantly to producing \ha\ emission
we can then compare our statistics to this mass value. Assuming a Salpeter IMF and an upper
mass limit for producing hydrogen rich CC SNe of 25\msun\ (i.e. the upper mass limit for red supergiants; 
\citealt{lev07}), we can calculate the range from 10\msun\ downwards (in progenitor
mass) that is consistent with our statistics of $\sim$35\%\ of SNII falling on sites of little or zero \ha\ emission. From
these assumptions we calculate a lower mass value for producing SNII (and also the IIP sub-type) of 7.8\msun , consistent
with that suggested by direct detections. Our results therefore seem to suggest that
a significant fraction of SNII are produced by progenitor stars of less than 10\msun .\\
JA06 discussed alternative explanations to the fact that we find a significant fraction of SNII
falling on sites of zero \ha\ flux. These assumed that CC SNe arise from stars of initial mass $>$10\msun .
Although as stated above there is growing evidence for the production of CC SNe from stars below
10\msun , the number of events used to make these constraints are still reasonably small and many stellar
evolution codes predict CC from stars only of 10\msun\ or higher (e.g. \citealt{rit99}).
Here we therefore summarise a number of other physical processes that may be at play in producing 
the excess of SNII that we find occurring away from sites of recent SF.\\
In JA06 we discussed the `runaway' hypothesis, that these SNe
did originally form
within an HII region but since moved to the position of the SN between stellar birth and death, due to some peculiar velocity.
Another possibility is the destruction of massive clusters before the epoch of SN. 
Recent observations and simulations 
\citep{good06,bast06} have shown that many massive stellar clusters will in fact be destroyed on
timescales of $\sim$10 Myr. 
Within the stellar cluster stars with the highest mass will explode as SNe first, thus 
exploding while the clusters are still stable and hence will be found to be associated with the \ha\ emission
produced from the ionisation of the local gas. These initial SNe (likely to be SNIc and Ib, see the next section) will drive the removal of gas from
the cluster eventually leading to its destruction. Therefore within this scenario there are two possible
processes that could lead to our result.
Firstly with gas removal from the system it may be that there is little gas to be ionised and therefore 
no host HII region will be seen at the site of some SNII SNe. Secondly, as the cluster is destroyed
while it attempts to regain virial equilibrium, many stars will be flung away with a high peculiar velocity leaving them 
far from their original host HII region.\\
Another explanation that was discussed in JA06 is the possibility that these SNe are occurring in regions
of dust content, through which the SNe are visible but the \ha\ emission is not. However, it is unclear why this would affect
the SNII much more than the SNIb/c. It has also been found, through mid-IR observations of the \textit{SINGS} survey,
that highly obscured SF regions only seen in the IR make up only a small ($\sim$4\%) fraction of the overall
SF distribution in nearby galaxies \citep{pres07}, arguing against this as a significant factor.\\

We conclude that the dominant effect producing our results on the association of SNII to 
the \ha\ emission of their host galaxies is that a significant fraction of SNII progenitors are stars with initial masses
below 10\msun . However, we also believe that it is likely that all the processes we discuss above play some part in producing the
observed NCR distribution.
We have discussed the various processes that could be involved
in shaping the results that we see, now we will explore how we can use these results to compare and constrain the
nature of the different progenitors of the different CC SN types.

\subsection{Relative progenitor masses}
\label{mass} 

\begin{figure}
\includegraphics[width=8.5cm]{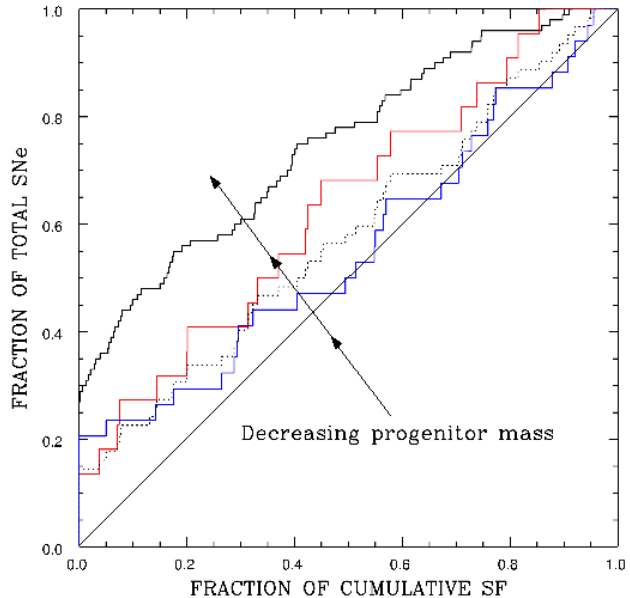}
\caption{Cumulative distribution of the NCR values for the various CC sub-types:  
black represents the overall SNII
population, red the SNIb, dotted the overall SNIb/c distribution and blue the SNIc. 
From comparing the distributions a sequence of decreasing progenitor mass emerges, indicated
on the plot from the SNIc to Ib and finally the II showing the lowest degree of association to the line emission. The 
solid black diagonal line shows the expected results for a hypothetical population that exactly traces the \ha\ line emission}
\label{cumuldist}
\end{figure}

From the arguments presented at the start of this section we can use comparisons of the mean NCR values
of the different SN types to compare the relative mass ranges of their progenitors.
The first conclusion is that 
we confirm the results of JA06 that overall the SNIb/c progenitor population arise from
more massive progenitors than the SNII. 
When we compare the CC SN sub-types in detail we find that the different CC sub-types appear to form a sequence of increasing 
progenitor mass, going
from SNII at the low mass end, through SNIb to SNIc as the highest mass progenitors. This sequence is illustrated in
Fig.~\ref{cumuldist}. In this figure we plot the cumulative distributions of the NCR values 
of the overall SNII, the SNIb/c, and the
SNIb and Ic distributions individually. We also plot a hypothetical distribution for a population that exactly
traces the line emission. The plot shows that the SNIc accurately traces the 
\ha\ emission, except for a slight excess of NCR values at zero. As we go to the other SN distributions we see
that they show an increasingly lower association to the line emission. A distinctive pattern emerges as indicated
by the arrows on the plot, going from high to lower and lower mass progenitors implied from the differences between
the distributions and the hypothetical flat distribution.
This sequence can be seen to fit to the paradigm where CC SNe (II, Ib then Ic) arise from stars of increasingly higher initial mass,
leading to stronger pre-SN stellar winds that strip the stars of their envelopes and produce
the observed differences we see in their spectra.\\ 

With the statistics presented in \S \ref{results} it is harder to make any firm statements as to differences within
the progenitor masses of the various SNII sub-types. However, with the small number of SNIIL as a strong caveat, it seems that
these SNe arise from similar mass progenitors to the SNIIP. This would imply that that metallicity or binarity may play a dominant role in deciding SN type, by enabling 
additional envelope stripping prior to explosion.\\
With respect to the SNIIb, although we only have 4 objects in our sample our results suggest that these
SNe arise from more massive progenitors than the overall SNII population. They also show a higher degree 
of association to the \ha\ emission than the SNIb (although again we stress the low statistics involved). 
A recent discovery of the progenitor of a IIb SN \citep{crock08} has suggested a possible progenitor mass of 28\msun , consistent with
our result that these SNe arise from towards the high end of the CC SN progenitor sequence. 
The only other direct detection of a SNIIb progenitor is that of SN 1993J. \cite{maun04} estimated that
this SN arose from a interacting binary with components of 14 and 15 \msun\ stars. Again our result that SNIIb 
arise from stars that follow the \ha\ emission of their host galaxies is consistent with this result.\\
One of the most interesting results to arise from this work regards the SNIIn. 
Our results (see \S ~\ref{IIn}) suggest that these SNe arise from similar mass 
progenitors to the overall SNII population and do not follow the underlying \ha\ emission of their host galaxies.
This would seem to be in conflict with recent thoughts on this SN type. Arguments 
have been put forward (e.g. \citealt{smith08} and references therein) that the observations of these SNe (strong
narrow emission lines and high luminosities) 
require high pre-SN mass loss rates and huge circumstellar envelopes, arising from 
only the most massive stars, which would presumably trace the \ha\ emission within galaxies. It has also been argued that two
SNIIn, (2005gl and 2005gj) had luminous blue variable (LBV) progenitors (\citealt{gal07,tru08}, respectively), again
stars of very high mass ($\sim$25-40\msun\ and above). Although some SNIIn probably do arise 
from very massive stars, our results suggest that the majority of these events arise from progenitors towards
the low end of the CC progenitor mass range.
A recent direct detection of the progenitor of the SNIIn 2008S on pre-explosion
\textit{SPITZER} mid-IR images \citep{pri08}, enabled an estimate to be made of the progenitor
mass of $\sim$10\msun\ consistent with our results (however, there is some debate as to whether
this is a true SN and it is unlike most other SNIIn; Smartt 2008, private communication). 
An intriguing possibility for progenitors from this mass range would be the super-AGB stars (SAGBs), a scenario
suggested by the modeling of \cite{po08}.
The initial mass range for SAGB evolution is 7.5-9.25\msun\ and it is thought that the upper mass
part of this range will produce electron-capture (EC) SNe \citep{po08}. The mass-loss rates of these 
systems can be extremely high due to a large number of thermal pulses, potentially producing the capacity 
for interaction of the
SN with a large amount of circumstellar material, and hence the narrow emission lines seen in SNIIn.\\
SN `impostors' are thought to be the outbursts of very massive unstable LBV stars \citep{van00,maun06}
that go through stages of intense mass loss, during which the luminosity of such objects can rise by 
more than three magnitudes (see \citealt{hump94} for a review on this subject), hence masquerading as `true' SNe. 
Given the presumed high mass nature of these events (and therefore their relatively short stellar lifetimes) one
would expect these events to trace the distribution of high mass SF within their host galaxies. However,
our results presented in \S ~\ref{imposres} would seem to be inconsistent with this picture. We find that
the SN `impostors' within our sample do not trace the underlying SF and in fact show the lowest degree of association of 
all SN types analysed in the current paper. We stress again here that there are only 6 such events within our sample 
and it is therefore hard to draw any firm conclusions before the statistics are improved. We note however that many LBVs observed in the local group
are often more isolated than one would expect and are not always found within dense young stellar clusters \citep{burg06}.

\section{Conclusions}
\label{con}
We find that there is a significant fraction of the SNII population that do not show any association to the distribution
of \ha\ line emission. This excess of $\sim$35\%\ of SNII falling on sites of
little or zero \ha\ flux, compared to what would be expected if
they accurately traced the underlying SF, suggests that a large fraction of SNII arise from progenitor stars of less than 
10\msun . 
Our results also 
imply that the different CC SN types can be separated into a sequence of increasing progenitor mass running from
the SNII through the Ib, with finally the SNIc arising from the highest mass progenitors. We now summarise our
findings on the possible relative mass ranges of the progenitors of the different CC SN types.
\begin{itemize}
\item
Assuming that only stars of 10\msun\ and above significantly contribute to the ionising flux that produces
\ha\ emission within galaxies, we calculate a lower mass limit for producing SNII of 7.8\msun .
\item
We confirm the results of JA06, that the SNIb/c trace the SF of
their host galaxies more accurately than the SNII, implying that they arise from a higher mass progenitor population than the SNII.
\item
SNIc accurately trace the underlying SF within their host galaxies and therefore probably arise from the highest mass
progenitors of all SNe.
\item
SNIIL show a similar degree of association to \ha\ emission as the overall SNII population implying that they arise 
from stars of similar mass to those of SNIIP, with metallicity or binarity probably playing
an important role in removing part of their envelopes and thus changing the shape of their light curves.
\item
Our results suggest that SNIIb arise from more massive stars than the overall SNII population.
\item
Although some SNIIn may arise from very massive stars, our results suggest that the majority come from the low end of the CC mass spectrum.
\item
SN `impostors' do not seem to trace the high mass SF within host galaxies. 
\end{itemize}


\section*{Acknowledgments}
We thank the referee, S. Smartt for his constructive comments that have
greatly improved the content of this paper.
We also thank Mike Irwin for processing the INT data through the CASU WFC automated reduction pipeline. This research
has made use of the NASA/IPAC Extragalactic Database (NED) which is operated by the Jet Propulsion Laboratory, California
Institute of Technology, under contract with the National Aeronautics and Space Administration. J. Eldridge, S. Percival and M. Salaris
are thanked for useful discussion and assistance.



\appendix

\section[]{Application of the Kolmogorov-Smirnov test to the SN data}
In this appendix we highlight a feature of commonly-used
implementations of the KS test, which caused particular problems for
the analysis presented in this paper.  These tests were implemented using
the on-line statistics calculator at
\vspace{2mm}\\
http://www.physics.csbsju.edu/stats/KS-test.html
\vspace{2mm}\\
but identical results were found with a direct implementation of the kstwo
code from `Numerical Recipes' \citep{press92}.\\
The problems were noted when we initially found apparently significant
differences between distributions of NCR values that to the eye
appeared quite similar. The $D$ statistic, parametrising the maximum difference
between pairs of normalised cumulative distributions, was for some
tests found to be significantly over-estimated.  It appears that this
occurs for those distributions with significant numbers of points with
identical values (which for our NCR distributions tend to be zeroes), 
and where the two samples are of different sizes.  The sceptical reader 
can quickly test this, using the above website, and the following points
as input:
\vspace{2mm}\\
0.01 0.23 0.32 0.40 0.40 0.40 0.40 0.51 0.59 0.63 0.67 0.73
\vspace{2mm}\\
Paste these numbers once into one of the data entry boxes, and twice
into the other, to give samples with identical normalised 
cumulative distributions, but different overall sizes.  This results
in an estimated $D$ of 0.1667, in spite of the identical cumulative 
distributions.  The overestimate of $D$ appears strongly dependent on the
number of identical points (large `steps' in the cumulative distribution),
which are a particular feature of our datasets, but will certainly
affect some other applications.\\
This does not appear to be a generally appreciated problem.  We advocate
careful checking of the $D$ value produced by KS software against an 
accurate plot of the normalised cumulative distributions, to ensure it is 
a real difference, and not an artefact caused by steps in the distributions.


\section[]{SN and host galaxy data}


%
%
%
%
%
%

\begin{table*}\label{SNtab} \centering
\caption{Data for all SNe and host galaxies}
\begin{tabular}[t]{cccccccc}
\hline
\hline
SN & Host galaxy &Galaxy type& V$_\textit{r}$ (\kms ) & SN type & NCR value & Telescope& Reference\\
\hline
1917A &  NGC 6946 &SABcd& 48  &  II &  0.207& INT& \\				       
1921B &  NGC 3184 &SABcd& 592 &  II &  0.000& INT& \\				       
1926A &  NGC 4303 &SABbc& 1566 & IIL &  0.078&INT& \\				       
1937F &  NGC 3184 &SABcd& 592 & IIP &  0.000& INT& \\				       
1940B &  NGC 4725 &SABab& 1206 & IIP &  0.000&INT& \\
1941A &  NGC 4559 &SABcd& 816 & IIL &  0.859& INT& \\				       				       
1941C &  NGC 4136 &SABc& 609 &  II &  0.000& JKT& \\				       
1948B &  NGC 6946 &SABcd& 48 & IIP &  0.387&  INT& \\
1954A &  NGC 4214 &IABm& 291 &  Ib &  0.000& INT& \\				       				       
1954C &  NGC 5879 &SAc& 772 &  II &  0.163& JKT& \\
1954J &  NGC 2403 &SABcd& 131 & `impostor'* & 0.187& INT &  \cite{vandyk05}\\	
1961I &  NGC 4303 &SABbc& 1566 &  II &  0.327&INT& \\				       			       
1961V &  NGC 1058 &SABc& 518 & `impostor'* & 0.363& JKT & \cite{good89}\\
1961U &  NGC 3938 &SABc& 809 & IIL &  0.000& LT& \\				       				       
1962L &  NGC 1073 &SABc& 1208 &  Ic &  0.000&JKT& \\
1964A &  NGC 3631 &SABc& 1156 &  II &  0.000&INT& \\
1964F &  NGC 4303 &SABbc& 1566 &  II &  0.000&INT& \\				       				       				       
1964H &  NGC 7292 &IBm& 986 &  II &  0.059& JKT& \\				       
1964L &  NGC 3938 &SABc& 809 &  Ic &  0.000& LT& \\				       
1965H &  NGC 4666 &SABc& 1529 & IIP &  0.597&LT& \\				       
1965N &  NGC 3074 &SABc& 5144 & IIP &  0.031&INT& \\				       
1965L &  NGC 3631 &SABc& 1156 & IIP &  0.001&INT& \\				       
1966B &  NGC 4688 &SBcd& 986 & IIL &  0.367& LT& \\				       
1966J &  NGC 3198 &SBc& 663 &  Ib &  0.000& INT& \\
1967H &  NGC 4254 &SAc& 2407 & II* & 0.568 & INT & \cite{vandyk92}\\
1968D &  NGC 6946 &SABcd& 48 &  II &  0.018&  INT& \\ 				       
1968I &  NGC 4254 &SAc& 2407 & IIP &  0.000&INT& \\						       
1968V &  NGC 2276 &SABc& 2410 &  II &  0.327&JKT& \\
1969B &  NGC 3556 &SBcd& 699 & IIP &  0.191& INT& \\				       				       		       
1969L &  NGC 1058 &SAc& 518 & IIP &  0.000& JKT& \\				       
1971S &  NGC 493  &SABcd& 2338 &IIP  &  0.174&JKT& \\				       
1971K &  NGC 3811 &SBcd& 3105 & IIP &  0.176&INT& \\				       
1972Q &  NGC 4254 &SAc& 2407 & IIP &  0.405&INT& \\				       
1972R &  NGC 2841 &SAb& 638 &  Ib &  0.071& INT& \\				       
1973R &  NGC 3627 &SABb& 727 & IIP &  0.325& INT& \\				       
1975T &  NGC 3756 &SABbc& 1318 & IIP &  0.000&INT& \\				       
1979C &  NGC 4321 &SABbc& 1571 & IIL &  0.000&LT& \\				       
1980K &  NGC 6946 &SABcd& 48 & IIL &  0.007&  INT& \\	
1982F &  NGC 4490 &SBd& 565 & IIP &  0.095&INT&\\				       			       
1983I &  NGC 4051 &SABbc& 700 &  Ic &  0.265&JKT&\\				       
1984E &  NGC 3169 &SAa& 1238 & IIL &  0.616&INT&\\
1985G &  NGC 4451 &Sbc& 864 & IIP &  0.641&INT&\\				       
1985F &  NGC 4618 &SBm& 544 & Ib* &  0.854& LT&\cite{gas86}\\					       
1985L &  NGC 5033 &SAc& 875 & IIL &  0.301&INT&\\
1987F &  NGC 4615 &Scd& 4716 & IIn &  0.352&INT&\\				       				       			       
1987K &  NGC 4651 &SAc& 805 & IIb &  0.746& JKT&\\				       
1987M &  NGC 2715 &SABc& 1339 &  Ic &  0.000&INT&\\				       
1988L &  NGC 5480 &SAc& 1856 &  Ib &  0.425&LT&\\				       
1989C &  UGC 5249 &SBd& 1874 & IIP &  0.689&LT&\\				       
1990E &  NGC 1035 &SAc& 1241 & IIP &  0.000&LT&\\				       
1990H &  NGC 3294 &SAc& 1586 &  IIP* &  0.000&INT& \cite{fil2_93}\\				       
1990U &  NGC 7479 &SBc& 2381 &  Ic &  0.712&JKT&\\
1991A &  IC 2973  &SBd& 3210 & Ic  &  0.773&INT&\\				       				       
1991G &  NGC 4088 &SABbc& 757 & IIP &  0.066&JKT& \\				       
1991N &  NGC 3310 &SABbc& 993 &  Ic &  0.759&JKT&\\				       
1992C &  NGC 3367 &SBc& 3040 &  II &  0.021&INT&\\
1993G &  NGC 3690 &Double system& 3121 &  IIL* &  0.064&INT& \cite{tsvet94}\\				       
1993X &  NGC 2276 &SABc& 2410 &  II &  0.039&JKT&\\				       
1994I &  NGC 5194 &SAbc & 463 &  Ic  & 0.550&INT&\\
1994Y &  NGC 5371 &SABbc & 2558& IIn  & 0.000&INT&\\			       		       
1994ak &  NGC 2782 &SABa & 2543&  IIn &  0.000&LT&\\
1995F  & NGC 2726  &SABc& 2410&  Ic  & 0.548&JKT&\\
1995N  & MCG -02-38-17 &IBm& 1856 & IIn & 0.001&LT&\\				       
				       
\hline
\end{tabular}
\end{table*}

\setcounter{table}{0}
\begin{table*} \centering
\caption{Data for all SNe and host galaxies}
\begin{tabular}[t]{cccccccc}
\hline
SN & Host galaxy & Galaxy type&V$_\textit{r}$ (\kms )& SN type & NCR value& Telescope & Reference\\
\hline
1995V &  NGC 1087  &SABc& 1517 &  II  & 0.424&JKT&\\				       
1995ag & UGC 11861& SABdm& 1481 &   II & 0.660&JKT&\\  				       
1996ae & NGC 5775 & Sb& 1681& IIn  & 0.747&JKT&\\				       
1996ak & NGC 5021 & SBb& 8487 &  II  & 0.562&INT&\\
1996aq & NGC 5584 & SABcd& 1638 &  Ic  & 0.050&LT&\\  				       
1996bu & NGC 3631 & SAc& 1156 & IIn  & 0.000&INT&\\	
1997bs & NGC 3627 & SABb& 727 & `impostor'*  & 0.023&INT& \cite{van00}\\   
1997X  & NGC 4691 & SBO/a& 1110 &  Ic  & 0.323&INT&\\				       			       
1997db & UGC 11861& SABdm& 1481 &   II & 0.029&JKT&\\				       
1997dn & NGC 3451 & Sd& 1334 &  II  & 0.073&JKT&\\
1997dq & NGC 3810 & SAc& 993 & Ic*  & 0.296&JKT& \cite{mazz04}\\ 				       
1997eg & NGC 5012 & SABc& 2619 & IIn  & 0.338&INT&\\				       
1997ei & NGC 3963 & SABbc& 3188 &  Ic  & 0.288&INT&\\				       
1998C  & UGC 3825 & SABbc& 8281 &  II  & 0.000&INT&\\
1998T  & NGC 3690 & Double system& 3121 &  Ib  & 0.578&INT&\\  				       
1998Y  & NGC 2415 & Im?& 3784 &  II  & 0.349&INT&\\				       
1998cc & NGC 5172 & SABbc& 4030 &  Ib  & 0.331&INT&\\	
1999D  & NGC 3690 & Double system& 3121 &  II  & 0.054&INT&\\  			       
1999br & NGC 4900 & SBd& 960 &  IIP*  & 0.099&JKT& \cite{ham03}\\
1999bu & NGC 3786 & SABa& 2678 &  Ic  & 0.000&INT&\\				       
1999bw & NGC 3198 & SBc& 663 & `impostor'*  & 0.000&INT& \cite{vandyk05}\\
1999dn & NGC 7714 & SBb& 2798 &  Ib  & 0.038&JKT&\\
1999ec & NGC 2207 & SABbc& 2741 &  Ib  & 0.815&INT&\\ 				       
1999ed & UGC 3555 & SABbc& 4835 &  II  & 0.615&INT&\\				       
1999em & NGC 1637 & SABc& 717 & IIP  & 0.394&LT&\\ 				       
1999gb & NGC 2532 & SABc& 5260 & IIn  & 0.676&INT&\\				       
1999gi & NGC 3184 & SABcd& 592 & IIP  & 0.637&INT&\\				       
1999gn & NGC 4303 & SABbc& 1566 &  IIP*  & 0.897&INT& \cite{past04}\\
2000C  & NGC 2415 & Im?& 3784 &  Ic  & 0.494&INT&\\
2000cr & NGC 5395 & SAb& 3491&  Ic  & 0.000&INT&\\				       
2000de & NGC 4384 & Sa& 2513 &  Ib  & 0.554&INT&\\	       				       
2000ew & NGC 3810 & SAc& 993 &  Ic  & 0.907&JKT&\\
2001B  & IC 391   & SAc& 1556 &   Ib & 0.201&INT&\\
2001M  & NGC 3240 & SABb& 3550 &  Ic  & 0.142&INT&\\				       
2001R  & NGC 5172 & SABbc& 4030 &  IIP*  & 0.000&INT& \cite{math3_01}\\      				       
2001aa & UGC 10888& SBb& 6149 &   II & 0.000&INT&\\				       
2001ac & NGC 3504 & SABab& 1534 & `impostor'*  & 0.000&INT& \cite{math01}\\
2001ai & NGC 5278 & SAb& 7541 &  Ic  & 0.878&INT&\\
2001co & NGC 5559 & SBb& 5166 & Ib/c  & 0.313&INT&\\				       
2001ef & IC 381   & SABbc& 2476 &   Ic & 0.944&INT&\\
2001ej & UGC 3829 & Sb& 4031 &  Ib  & 0.314&INT&\\   	       				       
2001fv & NGC 3512 & SABc& 1376 &  IIP*  & 0.169&INT& \cite{math2_01}\\				       
2001gd & NGC 5033 & SAc& 875 & IIb  & 0.459&INT&\\				       
2001is & NGC 1961 & SABc& 3934 &  Ib  & 0.449&INT&\\   				       
2002A  & UGC 3804 & SABbc& 2887 & IIn  & 0.401&JKT&\\
2002bm & MCG -01-32-19 &SBbc& 5462 &  Ic & 0.565&INT&\\				       
2002bu & NGC 4242 & SABdm& 506 & IIn  & 0.000&JKT&\\
2002ce & NGC 2604 & SBcd& 2078 &  II  & 0.108&JKT&\\
2002cg & UGC 10415& SABb& 9574 &   Ic & 0.955&INT&\\
2002cp & NGC 3074 & SABc& 5144 & Ib/c  & 0.131&INT&\\     				       
2002cw & NGC 6700 & SBc& 4588 &  Ib  & 0.370&INT&\\	
2002dw & UGC 11376& S& 6528 &   II & 0.475&INT&\\
2002ed & NGC 5468 & SABcd& 2842 & IIP  & 0.395&INT&\\
2002ei & MCG -01-09-24 &Sab& 2319 & IIP & 0.909&LT&\\				       
2002fj & NGC 2596 & Sb& 5938 & IIn  & 0.558&INT&\\  				       
2002gd & NGC 7537 & SAbc& 2674 &  II  & 0.167&JKT&\\
2002hh & NGC 6946 & SABcd& 48 & IIP  & 0.000&INT&\\
2002hn & NGC 2532 & SABc& 5260 &  Ic  & 0.672&INT&\\
2002ho & NGC 4210 & SBb& 2732 &  Ic  & 0.405&INT&\\  
2002ji & NGC 3655 & SAc& 1473& Ib/c  & 0.078&INT&\\				       
2002jz & UGC 2984 & SBdm& 1543 &  Ic  & 0.513&INT&\\   
2002kg & NGC 2403 & SABcd& 131 & `impostor'*  & 0.055&INT& \cite{maun06}\\
2003H  & NGC 2207 & SABbc& 2741 &  Ib  & 0.144&INT&\\	 
\hline
\end{tabular}
\end{table*}

\setcounter{table}{0}
\begin{table*} \centering
\caption{Data for all SNe and host galaxies}
\begin{tabular}[t]{cccccccc}
\hline
SN & Host galaxy & Galaxy type &V$_\textit{r}$ (\kms ) & SN type & NCR value& Telescope & Reference\\
\hline				       
2003T  & UGC 4864 &SAab&  8368 &  II &  0.056&INT&\\   		       	       
2003Z  & NGC 2742 &SAc&  1289 &  IIP* &  0.013&JKT& \cite{past04}\\
2003ab & UGC 4930 &Scd&  8750 &  II &  0.000&INT&\\
2003ao & NGC 2993 &Sa&  2430 & IIP &  0.157&LT&\\			       
2003at & MCG +11-20-23 &Sbc  & 7195 &  II &  0.728&INT&\\
2003bp & NGC 2596  &Sb& 5938 &  Ib &  0.075&INT&\\
2003db & MCG +05-23-21 & S? & 8113 &  II &  0.150&INT&\\
2003ed & NGC 5303 &Pec & 1419 & IIb &  0.554&LT&\\
2003ef & NGC 4708 & SAab &4166 & II* & 0.257 & INT & \cite{gan03}\\
2003el & NGC 5000 &SBbc & 5608 &  Ic &  0.728&INT&\\				       
2003hp & UGC 10942& SB& 6378 &   Ic&  0.000&INT&\\		       
2003hr & NGC 2551 & SAO/a& 2344 &  II &  0.000&JKT&\\				       
2003ie & NGC 4051  & SABbc& 700 &  II &  0.373&JKT&\\		       
2003ig & UGC 2971  & S& 5881 &  Ic &  0.769&INT&\\				       
2004A  & NGC 6207  & SAc& 852 &  IIP* &  0.000&JKT& \cite{hen06}\\
2004C  & NGC 3683  & SBc& 1716 &  Ic &  0.920&INT&\\
2004G  & NGC 5668  & SAd& 1582 &  II &  0.000&INT&\\
2004ao & UGC 10862 & SBc& 1691 &   Ib&  0.420&INT&\\				       
2004bm & NGC 3437  & SABc& 1283 &  Ic &  0.704&INT&\\
2004bs & NGC 3323  & SB?& 5164 &  Ib &  0.200&INT&\\
2004dg & NGC 5806  & SABb& 1359 &  IIP* &  0.554&JKT& S. Smartt (2008, priv comm)\\
2004dk & NGC 6118  & SAcd& 1573 &  Ib &  0.794&INT&\\				       
2004ep &  IC 2152  & SABab& 1875 &  II &  0.289&LT&\\	       			       
2004gq & NGC 1832  & SBbc& 1939 &  Ib &  0.738&LT&\\
2004gt & NGC 4038  & SBm& 1642 & Ib/c &  0.758&LT&\\	       				       
2004ge & UGC 3555  & SABbc& 4835 &  Ic &  0.293&INT&\\
2005O  & NGC 3340  & S& 5558 &  Ib &  0.709&INT&\\	       
2005V  & NGC 2146  & SBab& 893 & Ib/c &  0.000&LT&\\
2005ad & NGC 941   & SABc& 1608 & IIP*  &  0.000&INT& S. Smartt (2008, priv comm)\\
2005ay & NGC 3938  & SAc& 809 & IIP &  0.873&LT&\\
2005az & NGC 4961  & SBcd& 2535 & Ic* &  0.000&LT& \cite{burk05}\\	       				       
2005cs & NGC 5194  & SAbc& 463 & IIP &  0.396&INT&\\	       				       
2005dl & NGC 2276  & SABc& 2410 &  II &  0.730&INT&\\				       
2005dp & NGC 5630  & Sdm& 2655 &  II &  0.511&LT&\\				       
2005kk & NGC 3323  & SB?& 5164 &  II &  0.116&INT&\\				       
2005kl & NGC 4369  & SAa& 1045 &  Ic &  0.570&LT&\\				       
2005lr & ESO 492-G2& SAb& 2590 &  Ic &  0.175&LT&\\ 				       
2006am & NGC 5630 & Sdm & 2655 & IIn &  0.000&LT&\\
2006gi & NGC 3147 & SAbc& 2820 &  Ib &  0.000&INT&\\				       
2006jc & UGC 4904 & SB& 1670 & Ib/c &  0.172&LT&\\	       				       
2006ov & NGC 4303 & SABbc& 1566 & IIP &  0.284&INT&\\
2008ax & NGC 4490 & SBd& 565 & IIb &  0.080&INT&\\					       
\hline
\end{tabular}

\caption{Data for all SNe and host galaxies: Columns 1 and 2 give the SN and host galaxy respectively. In columns 3 and 4 we
present the morphological type and recession velocities of the host galaxies (both taken from NED). In column 5 the SN types are listed and the NCR data
for each SN are given in column 6.
In column 7 the telescope used for imaging of the SN host galaxy is given and for SNe where type classification was changed from those given in the 
Asiago catalogue a reference for the new designated type is given in the final column, and these type classifications are marked with an asterisk.}
\end{table*}


\label{lastpage}	    
			    
\end{document}